%% file: pulling_brillouin1.tex
\documentclass[journal]{IEEEtran}
\usepackage{cite}
\usepackage{graphicx}
\usepackage{psfrag}

\begin{document}
\title{Polarization dependent Brillouin gain in randomly birefringent fibers}
\author{
Leonora Ursini,~\IEEEmembership{Member,~IEEE} Marco Santagiustina,~\IEEEmembership{Member,~IEEE} Luca Palmieri,~\IEEEmembership{Member,~IEEE}
\thanks{Manuscript received \ldots}
\thanks{The authors are with the Department of Information Engineering, University of
Padova, 35131 Padova, Italy (e-mail: ursinile@dei.unipd.it). The research leading to these results has received funding from the European
Community's Seventh Framework Programme under grant agreement No. 219299, Gospel. Partial support from the Italian Ministry of Foreign Affairs
(Direzione Generale per la Promozione e la Cooperazione Culturale) and from the University of Padova (project ``Fiber optic polarimetric sensors
for extra high voltage transmission lines monitoring'') is acknowledged. This research was held in the framework of the agreement with ISCOM
(Rome).}}
\markboth{Photonics Technology Letters}{Shell \MakeLowercase{\textit{et al.}}: Bare Demo of IEEEtran.cls for Journals}
\maketitle
\begin{abstract}
An extensive study of the alignment between the pump, the signal and the polarization dependent gain (PDG) vectors in stimulated Brillouin
amplification in randomly birefringent fibers is realized by numerically integrating the equations governing the propagation. At the fiber
output, the signal tends to align to the PDG vector for large pump power because of the nonlinear polarization pulling effect. The PDG vector,
for large random birefringence, aligns to a state that has the same linear component of the pump but opposite circular component.
\end{abstract}
\section{Introduction}
\IEEEPARstart{O}{ver} the years, stimulated Brillouin scattering (SBS) in optical fibers has been investigated and
different applications have been found. For instance: optical signal amplification \cite{OlssonAPL1986},
distributed sensing \cite{BaoOL1993}, microwave signal generation \cite{YaoJOSAB1996}, tunable
delay lines \cite{SongOE2005}, microwave-photonics filter design \cite{SaguesPTL2007} and slow-fast light
generation \cite{sflbook2008}.
\IEEEpubidadjcol
SBS amplification efficiency in polarization-maintaining fibers depends on the pump and signal relative state of polarization (SOP)
\cite{StolenJQE1979}: SBS gain is maximum (zero) for parallel (orthogonal) pump-signal SOPs. In real fibers the input SOPs are not preserved
because of the random birefringence (polarization-mode dispersion - PMD), \cite{vanDeventerJLT1994}. The typical way to eliminate such effect is
by scrambling the pump polarization; however, this causes a severe reduction of the gain \cite{StolenJQE1979}. Anyway, recently, interesting
applications of polarized SBS have been pointed out; in particular, the nonlinear polarization pulling (NLPP) effect for the synthesis of
arbitrary polarization states has been first mentioned in \cite{ThevenazOFC2008}. In an analysis of SBS gain in randomly birefringent and spun
fibers \cite{GaltarossaPTL2008}, NLPP has been found in the solutions, but not addressed in detail. A more focused analysis on NLLP has been
presented in \cite{ZadokOE2008}, where it was pointed out that NLPP consists in the attraction of the signal SOP towards the direction of maximum
gain. In this Letter, the NLPP analysis is improved by showing how the polarized amplification can be elegantly formalized in terms of the
Polarization Dependent Gain (PDG) vector. This formulation provides a useful modeling tool to understand the physical limits of the NLPP
mechanism for arbitrary SOP synthesis. In particular, by extensive simulations, the importance of the regime of transition from low to large PMD
in determining the quality of the signal SOP will be underlined.

\section{Model}
In the undepleted pump approximation and by neglecting all nonlinear effects except SBS,
the evolution of the Stokes vectors of a forward-propagating signal $\overline{S}(z)=S_0(z)\hat{s}(z)$ and of a
counter-propagating pump $\overline{P}(z)=P_0(z)\hat{p}(z)$ is given by \cite{GaltarossaPTL2008},
\begin{eqnarray}
  \label{eqP} \frac{d\overline{P}}{dz} &=& \alpha_p \overline{P}-\mathbf{M}\omega_p \bar{b}\times \overline{P}, \\
  \label{eqS} \frac{d\overline{S}}{dz} &=& -\alpha_s \overline{S}+\frac{g}{2}[P_0\overline{S}+S_0\overline{P}]+\omega_s
    \bar{b}\times \overline{S}.
\end{eqnarray}
The parameters are defined as follows: attenuation coefficient $\alpha_s=\alpha_p=0.2\;dB/km$; SBS gain coefficient
$g=0.625\;W^{-1}m^{-1}$; signal and pump angular frequency $\omega_s=2\pi c/\lambda_s$ ($c=3\cdot 10^8\;m/s$,
$\lambda_s=1550\;nm$) and $\omega_p=\omega_s+\Omega_B$ ($\Omega_B=2\pi \Delta f_B$, $\Delta f_B=11.25\;GHz$ is the
Brillouin frequency shift). The matrix $\mathbf{M}=diag(1,1,-1)$ accounts for signal and pump counter-propagation
\cite{GaltarossaPTL2008,GaltarossaJLT2006}. The random birefringence vector $\omega_{p,s}\bar{b}$ is
obtained through the random modulus model (RMM) \cite{WaiJLT1996}. Let us remark that the RMM describes PMD through
two main parameters, the beat length $L_B$, which depends on the angular frequency, and the birefringence correlation
length $L_F$, that here was fixed to $L_F=10\;m$.
Both quantities contribute to determine the PMD coefficient \cite{WaiJLT1996,GaltarossaJLT2006}, hereinafter defined as
$D=\sqrt{\langle \Delta \tau^2 \rangle / L}$, where $\langle \Delta \tau^2 \rangle$ is the fiber mean square differential
group delay and $L=2\;km$ is the fiber length.

The PDG vector is defined in Stokes formalism as the vector $\overline{\Gamma}=\Gamma \hat{\Gamma}$ whose direction $\hat{\Gamma}$ is parallel to
the direction of the signal experiencing the maximum gain and whose modulus $\Gamma$ is such that the PDG in decibels (i.e. the difference
between the maximum and the minimum achievable gain) reads $PDG=10\log_{10}\left[(1+\Gamma)/(1-\Gamma)\right]$
\cite{HuttnerJSTQE2000,GaltarossaPTL2003b}.

The equation governing the evolution of $\overline{\Gamma}$ can be obtained from eq. \ref{eqS}, written in
Jones formalism \cite{HuttnerJSTQE2000}:
\begin{equation} \label{eqSjones}
\frac{d| A_s \rangle}{dz} =\left [ \left ( -\frac{\alpha}{2}+\frac{g}{4}P_0 \right )  + \left (
\frac{g}{4}\overline{P}-\frac{j}{2}\omega_s\bar{b} \right )\cdot \mathbf{\sigma} \right ]| A_s \rangle,
\end{equation}
where $| A_s \rangle$ is the signal Jones signal vector related to the corresponding Stokes vector by:
$\overline{S}=\langle A_s| \mathbf{\sigma} | A_s \rangle$, where $\mathbf{\sigma}$ is the vector
of Pauli matrices \cite{GordonPNAS2000}.
By following ref. \cite{HuttnerJSTQE2000}, starting from eq.
\ref{eqSjones}, the evolution equation of $\overline{\Gamma}$ can be straightforwardly determined, and reads:
\begin{equation}
\label{eqPDG} \frac{d\overline{\Gamma}}{dz} = \frac{g}{2}\overline{P}-\frac{g}{2}(\overline{P}\cdot\overline{\Gamma})
\overline{\Gamma}+\omega_s \bar{b}\times \bar{\Gamma}.
\end{equation}
%
%
\begin{figure}[!t]
\centering
\input{sPDG.tex} \caption{a) Mean and b) STD of the angle between the signal SOP $\hat{S}$ and the PDG versor
$\hat{\Gamma}$ at $z=L$, as a function of the PMD coefficient $D$. The pump input SOP is linear. Circles, diamonds and squares refer to
$P_0(L)=2,9,18\;mW$ respectively.}\label{s-PDG}
\end{figure}
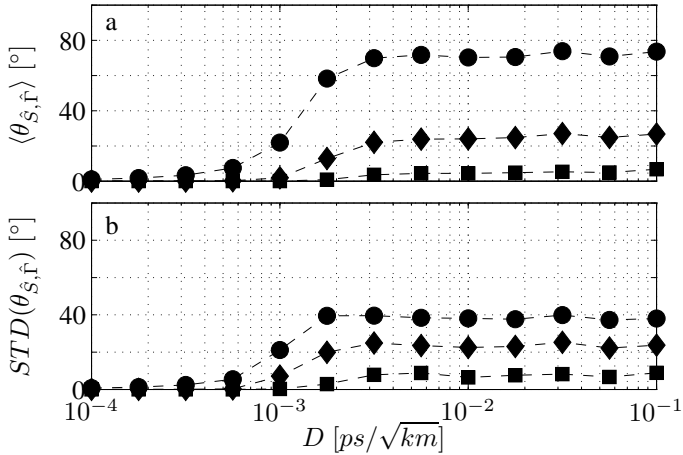
\section{Analysis}
Eqs. \ref{eqP}-\ref{eqS} and \ref{eqPDG} have been numerically integrated over a statistical ensemble of 4000 fiber birefringence realizations.
The condition $P_0(L)<18mW$ is used to maintain valid the undepleted pump approximation. The pump and signal input SOPs are fixed to the same
polarization at $z=L$ and $z=0$ respectively.

The NLPP effect, as a function of the PMD coefficient $D$ and of the input pump power $P_0(L)$, is demonstrated in Fig. \ref{s-PDG}, where the
mean and the standard deviation (STD) of the angle between the signal SOP and the PDG vector at the fiber output ($z=L$) is shown. Only the
linear pump input SOP is presented, the other pump input SOPs showing a similar behavior. Three values of the pump input power are presented:
circles, diamonds and squares refer to $P_0(L)=2,9,18\;mW$ respectively. Note that, taking the limit $z \rightarrow 0$ in eq. \ref{eqPDG}, the
direction of the maximum gain ($\hat{\Gamma}$) tends to coincide with the pump SOP at $z=0$.

In the low PMD regime ($D<10^{-4}\;ps/\sqrt{km}$, not represented in the figures),
the output signal SOP $\hat{S}$ is aligned to $\hat{\Gamma}$, with
negligible fluctuations, because all the SOPs and the direction of $\hat{\Gamma}$ are preserved during the propagation;
actually, the fiber is isotropic.
In the high PMD regime ($D>10^{-2}\;ps/\sqrt{km}$) the SOPs are highly scrambled; if the pump power is low,
the relative alignment is lost and the fluctuations are very large (circles).
However, if the pump power grows, the output signal is pulled toward the PDG vector,
with an increasingly reducing STD (diamonds and squares).
These results confirm the analysis of NLPP conducted in \cite{ZadokOE2008}.
\begin{figure}[!t]
\centering
\input{DOP.tex} \caption{Signal degree of polarization at $z=L$, as a function of the PMD coefficient $D$. The pump
input SOP is linear. Circles, diamonds and squares refer to $P_(L)=2,9,18\;mW$ respectively.}\label{DOPvsD}
\end{figure}
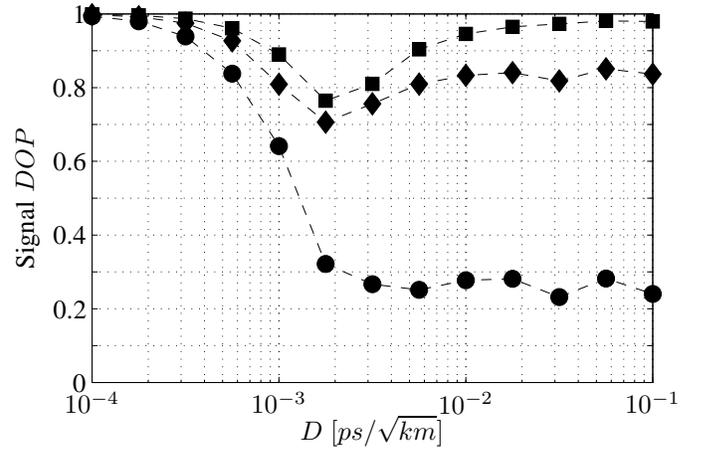

In order to evaluate the quality of the signal SOP exiting a device based on SBS-NLPP, the output signal degree of polarization
$DOP=\sqrt{\langle s_1^2 \rangle+\langle s_2^2 \rangle+\langle s_3^2 \rangle}$, where $\langle s_i^2 \rangle$, i=1,2,3 are the output signal mean
squared Stokes versor components, has been calculated. In Fig. \ref{DOPvsD} the $DOP$ is shown as a function of $D$, for a linear pump input SOP,
the other pump input SOPs showing a similar behavior. When the PMD influence is negligible $DOP \rightarrow 1$. In the high PMD regime, the $DOP$
is much less than 1, when the pump power is low, as previously predicted \cite{vanDeventerJLT1994}, however it increases by increasing the pump
power, owing to the NLPP effect. Remarkably, in the regime of transition from low to high PMD even for large pump powers, i.e. when the NLPP is
already very strong (squares in fig. \ref{s-PDG}), the signal $DOP$ is appreciably less than 1.

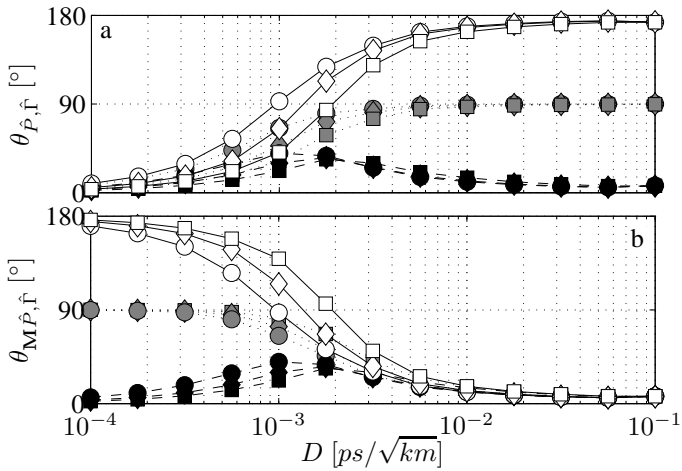
\begin{figure}[!t]
\centering
\input{pPDG_Mp_PDG.tex} \caption{a) Mean angle between the pump SOP $\hat{P}$ and $\hat{\Gamma}$ at $z=L$,
as a function of the PMD coefficient $D$. The black, grey, white markers with dashed, dotted, continuous curves refer to linear, elliptical,
circular input pump SOP. Circles, diamonds and squares refer to $P_0(L)=2,9,18\;mW$ respectively. b) Mean angle between $\mathbf{M} \hat{P}$ and
$\hat{\Gamma}$ at $z=L$, as a function of the PMD coefficient $D$. The symbols are the same as in a).}\label{p-Mp-PDG}
\end{figure}

To investigate further this limitation in the signal SOP quality, the relation between the PDG vector and the pump is studied. In Fig.
\ref{p-Mp-PDG}a, the mean alignment between the pump SOP and the PDG vector at $z=L$ is presented, as a function of $D$, for three different pump
input SOPs: linear ($\overline{P}(L)=(1,0,0)^T$ - black markers and dashed curves), elliptical ($\overline{P}(L)=(1/\sqrt(2),0,1/\sqrt(2)^T$ -
grey markers and dotted curves), righthand circular ($\overline{P}(L)=(0,0,1)^T$ - white markers and continuous curves).
Circles, diamonds and squares refer to $P_0(L)=2,9,18\;mW$ respectively. In the regime in which the PMD effect is negligible, the initial
alignment is preserved during the propagation. In the high PMD regime, the PDG vector does not tend to align to the pump SOPs but to
$\mathbf{M}\hat{P}$, which corresponds to a SOP with the same linear component of $\hat{P}$, but with the opposite circular component. This fact,
also shown in \cite{ZadokOE2008}, is confirmed in Fig. \ref{p-Mp-PDG}b, which presents the mean alignment between $\mathbf{M}\hat{P}$ and
$\hat{\Gamma}$, at $z=L$, as a function of $D$. However, in between the low and high PMD regimes studied in \cite{ZadokOE2008} there exists a
transition regime in which the output pump-PDG vector alignment strongly depends on the input SOP, on the pump power, and moreover presents large
fluctuations.
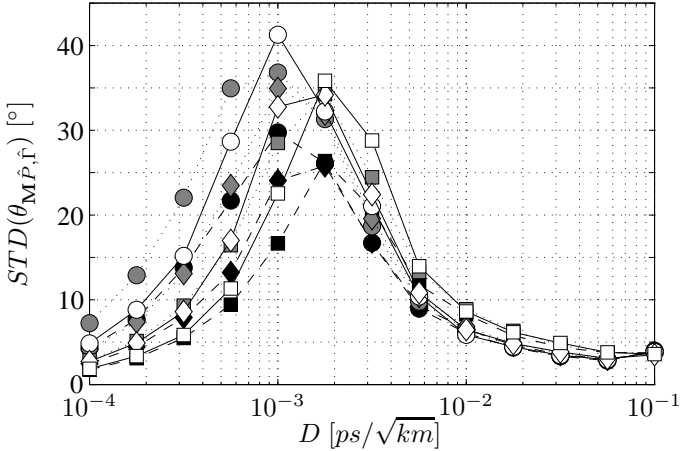
\begin{figure}[!t]
\centering
\input{std_MpPDG.tex} \caption{Standard deviation of angle between $\mathbf{M} \hat{P}$ and the PDG versor $\hat{\Gamma}$
at $z=L$, as a function of the PMD coefficient $D$. The black, grey, white markers with dashed, dotted, continuous curves
refer to linear, elliptical, circular input pump SOP. Circles, diamonds and squares refer to $P_0(L)=2,9,18\;mW$
respectively. }\label{std-MpPDG}
\end{figure}

In fact, in the transition regime the STD of the alignment between $\hat{\Gamma}$ and $\mathbf{M}\hat{P}$, at $z=L$ can be as large as 40
degrees, as shown in Fig. \ref{std-MpPDG}. The STD is also strongly dependent on the input SOP and on the pump power. To summarize, in the
transition regime, the signal SOP is strongly pulled towards the PDG vector (see fig. \ref{s-PDG}), however the PDG vector has not yet converged
to a stable, predictable configuration. Paradoxically, to improve the control and the quality of the synthesized output SOP, fibers with large
randomness are to be used.

\section{Conclusions}
In conclusion, the polarizing properties of stimulated Brillouin scattering have been studied over a large range of
PMD coefficient values in order to assess the quality and reliability of this
effect for the synthesis of arbitrary states of polarization. Three regimes have been identified.

For negligible polarization mode dispersion, the polarization dependent gain vector
is aligned with pump and the signal; the signal is fully polarized.

For high polarization mode dispersion, the polarization dependent gain vector tends to align to a state of polarization which has the same linear
component of the pump but opposite circular component. In this regime the nonlinear polarization pulling effect, i.e. the attraction of the
signal to the polarization dependent gain vector, is observed as the pump power increases. The signal is not fully polarized, though the degree
of polarization tends to 1 as the pump power increases.

In the transition regime the alignment among the vectors (PDG, signal and pump) is highly stochastic; it also
depends on the signal input polarization state and on the pump power; the degree of polarization is less than one.

Paradoxically, large polarization mode dispersion increases the quality and robustness of the polarization
control.

\end{document}

%% file: sPDG.tex
%
%
\begin{psfrags}%
\psfragscanon%
%
\psfrag{s01}[b][b]{$\langle \theta_{\hat{S},\hat{\Gamma}} \rangle\;[^{\circ}]$}%
\psfrag{s05}[t][t]{$D\;[ps/\sqrt{km}]$}%
\psfrag{s06}[b][b]{$STD(\theta_{\hat{S},\hat{\Gamma}})\;[^{\circ}]$}%
\psfrag{s17}[l][l]{b}%
\psfrag{s18}[lt][lt]{a}%
\psfrag{s19}[lt][lt]{b}%
%
\psfrag{x01}[t][t]{0}%
\psfrag{x02}[t][t]{0.1}%
\psfrag{x03}[t][t]{0.2}%
\psfrag{x04}[t][t]{0.3}%
\psfrag{x05}[t][t]{0.4}%
\psfrag{x06}[t][t]{0.5}%
\psfrag{x07}[t][t]{0.6}%
\psfrag{x08}[t][t]{0.7}%
\psfrag{x09}[t][t]{0.8}%
\psfrag{x10}[t][t]{0.9}%
\psfrag{x11}[t][t]{1}%
\psfrag{x12}[t][t]{$10^{-4}$}%
\psfrag{x13}[t][t]{$10^{-3}$}%
\psfrag{x14}[t][t]{$10^{-2}$}%
\psfrag{x15}[t][t]{$10^{-1}$}%
\psfrag{x16}[t][t]{}%
\psfrag{x17}[t][t]{}%
\psfrag{x18}[t][t]{}%
\psfrag{x19}[t][t]{}%
%
\psfrag{v01}[r][r]{0}%
\psfrag{v02}[r][r]{0.1}%
\psfrag{v03}[r][r]{0.2}%
\psfrag{v04}[r][r]{0.3}%
\psfrag{v05}[r][r]{0.4}%
\psfrag{v06}[r][r]{0.5}%
\psfrag{v07}[r][r]{0.6}%
\psfrag{v08}[r][r]{0.7}%
\psfrag{v09}[r][r]{0.8}%
\psfrag{v10}[r][r]{0.9}%
\psfrag{v11}[r][r]{1}%
\psfrag{v12}[r][r]{0}%
\psfrag{v13}[r][r]{}%
\psfrag{v14}[r][r]{40}%
\psfrag{v15}[r][r]{}%
\psfrag{v16}[r][r]{80}%
\psfrag{v17}[r][r]{}%
\psfrag{v18}[r][r]{0}%
\psfrag{v19}[r][r]{}%
\psfrag{v20}[r][r]{40}%
\psfrag{v21}[r][r]{}%
\psfrag{v22}[r][r]{80}%
\psfrag{v23}[r][r]{}%
%
\includegraphics[width=\columnwidth]{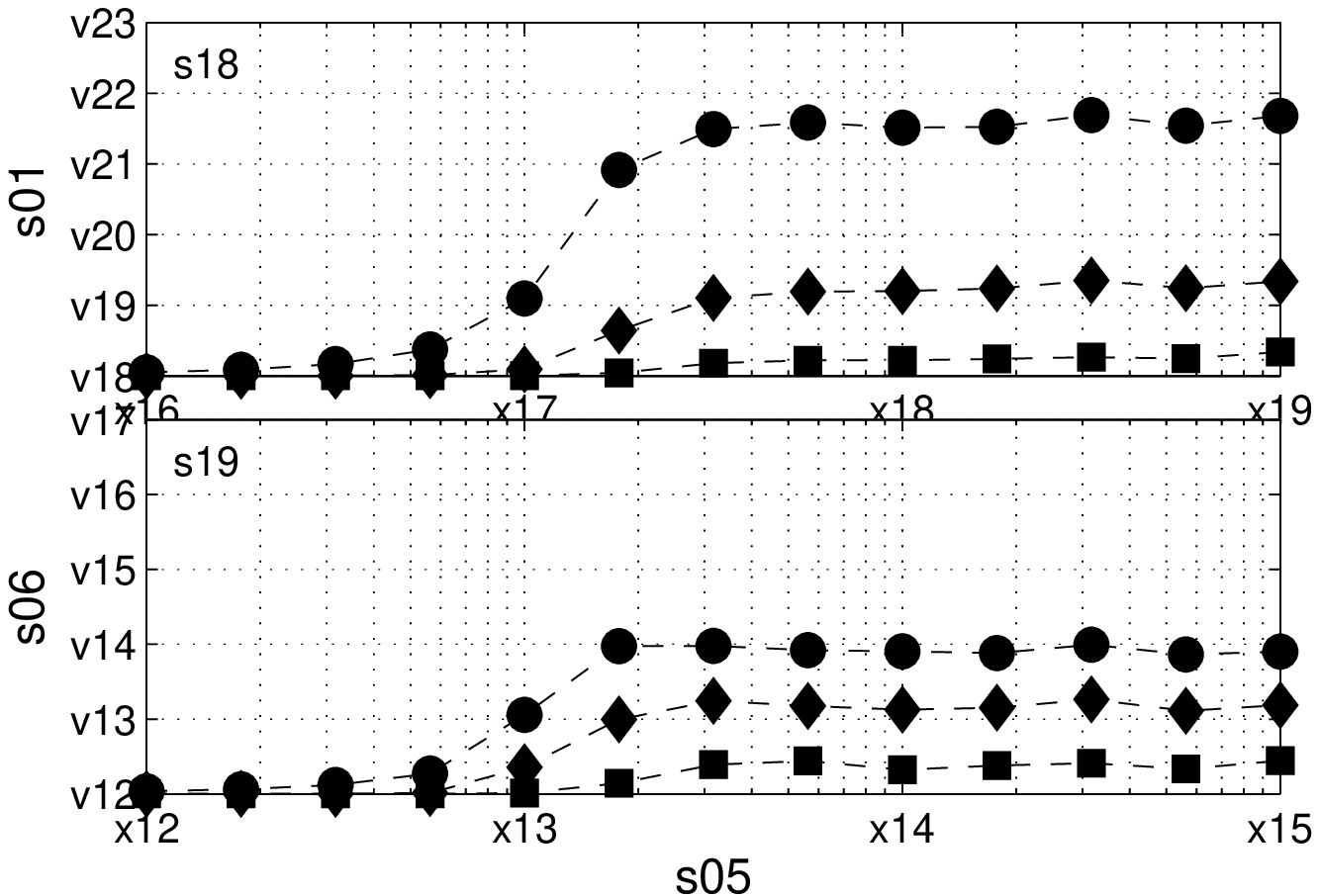}%
\end{psfrags}%
%

%% file: DOP.tex
%
%
\begin{psfrags}%
\psfragscanon%
%
\psfrag{s05}[t][t]{$D\;[ps/\sqrt{km}]$}%
\psfrag{s06}[b][b]{Signal $DOP$}%
%
\psfrag{x01}[t][t]{0}%
\psfrag{x02}[t][t]{}%
\psfrag{x03}[t][t]{0.2}%
\psfrag{x04}[t][t]{}%
\psfrag{x05}[t][t]{0.4}%
\psfrag{x06}[t][t]{}%
\psfrag{x07}[t][t]{0.6}%
\psfrag{x08}[t][t]{}%
\psfrag{x09}[t][t]{0.8}%
\psfrag{x10}[t][t]{}%
\psfrag{x11}[t][t]{1}%
\psfrag{x12}[t][t]{$10^{-4}$}%
\psfrag{x13}[t][t]{$10^{-3}$}%
\psfrag{x14}[t][t]{$10^{-2}$}%
\psfrag{x15}[t][t]{$10^{-1}$}%
%
\psfrag{v01}[r][r]{0}%
\psfrag{v02}[r][r]{}%
\psfrag{v03}[r][r]{0.2}%
\psfrag{v04}[r][r]{}%
\psfrag{v05}[r][r]{0.4}%
\psfrag{v06}[r][r]{}%
\psfrag{v07}[r][r]{0.6}%
\psfrag{v08}[r][r]{}%
\psfrag{v09}[r][r]{0.8}%
\psfrag{v10}[r][r]{}%
\psfrag{v11}[r][r]{1}%
\psfrag{v12}[r][r]{0}%
\psfrag{v13}[r][r]{}%
\psfrag{v14}[r][r]{0.2}%
\psfrag{v15}[r][r]{}%
\psfrag{v16}[r][r]{0.4}%
\psfrag{v17}[r][r]{}%
\psfrag{v18}[r][r]{0.6}%
\psfrag{v19}[r][r]{}%
\psfrag{v20}[r][r]{0.8}%
\psfrag{v21}[r][r]{}%
\psfrag{v22}[r][r]{1}%
%
\includegraphics[width=\columnwidth]{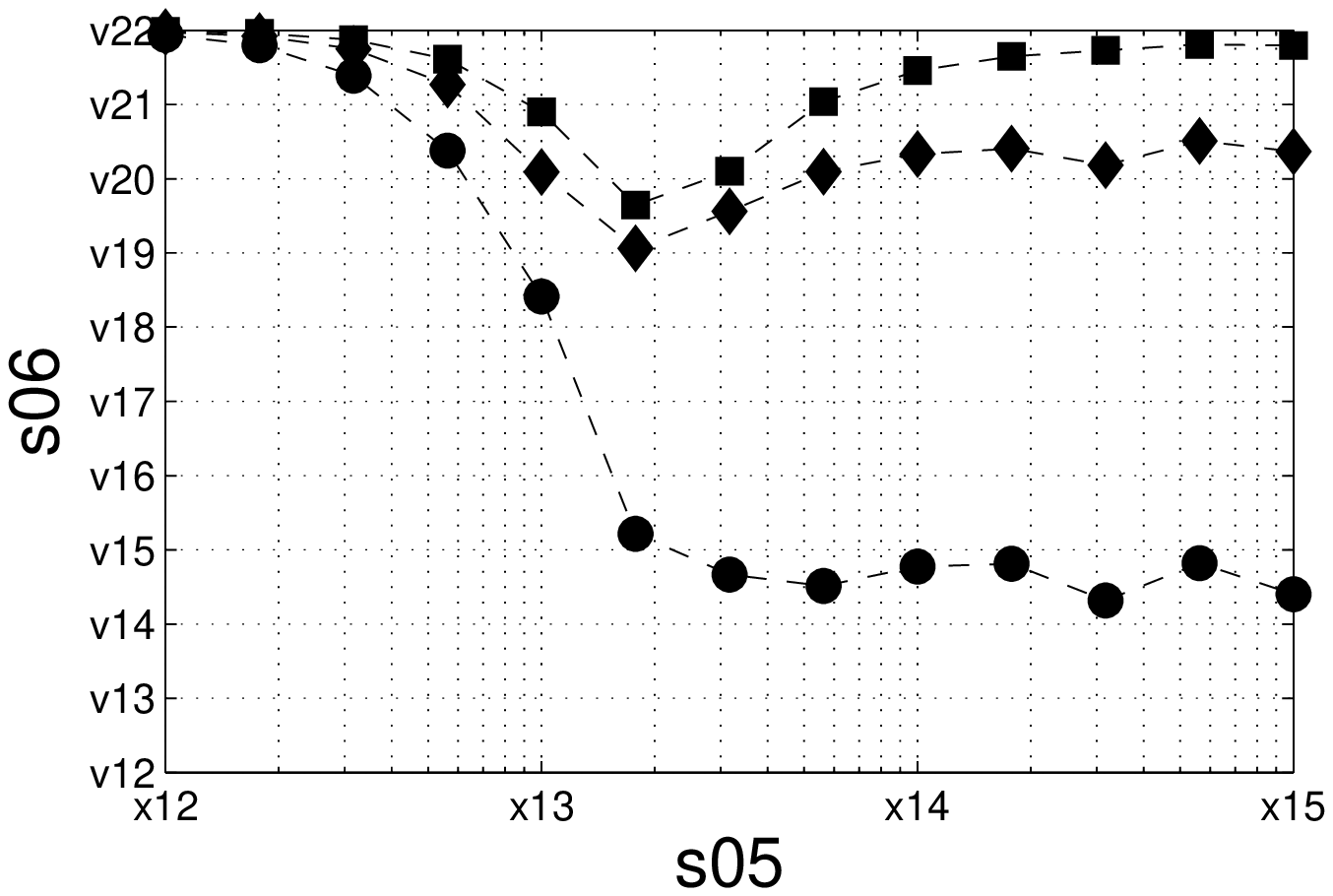}%
\end{psfrags}%
%

%% file: pPDG_Mp_PDG.tex
%
%
\begin{psfrags}%
\psfragscanon%
%
\psfrag{s02}[b][b]{$\theta_{\hat{P},\hat{\Gamma}}\;[^{\circ}]$}%
\psfrag{s05}[t][t]{$D\;[ps/\sqrt{km}]$}%
\psfrag{s06}[b][b]{$\theta_{\mathbf{M}\hat{P},\hat{\Gamma}}\;[^{\circ}]$}%
\psfrag{s21}[lt][lt]{a}%
\psfrag{s22}[l][l]{b}%
\psfrag{s23}[lt][lt]{b}%
%
\psfrag{x01}[t][t]{0}%
\psfrag{x02}[t][t]{0.1}%
\psfrag{x03}[t][t]{0.2}%
\psfrag{x04}[t][t]{0.3}%
\psfrag{x05}[t][t]{0.4}%
\psfrag{x06}[t][t]{0.5}%
\psfrag{x07}[t][t]{0.6}%
\psfrag{x08}[t][t]{0.7}%
\psfrag{x09}[t][t]{0.8}%
\psfrag{x10}[t][t]{0.9}%
\psfrag{x11}[t][t]{1}%
\psfrag{x12}[t][t]{$10^{-4}$}%
\psfrag{x13}[t][t]{$10^{-3}$}%
\psfrag{x14}[t][t]{$10^{-2}$}%
\psfrag{x15}[t][t]{$10^{-1}$}%
\psfrag{x16}[t][t]{}%
\psfrag{x17}[t][t]{}%
\psfrag{x18}[t][t]{}%
\psfrag{x19}[t][t]{}%
\psfrag{x20}[t][t]{}%
\psfrag{x21}[t][t]{}%
%
\psfrag{v01}[r][r]{0}%
\psfrag{v02}[r][r]{0.1}%
\psfrag{v03}[r][r]{0.2}%
\psfrag{v04}[r][r]{0.3}%
\psfrag{v05}[r][r]{0.4}%
\psfrag{v06}[r][r]{0.5}%
\psfrag{v07}[r][r]{0.6}%
\psfrag{v08}[r][r]{0.7}%
\psfrag{v09}[r][r]{0.8}%
\psfrag{v10}[r][r]{0.9}%
\psfrag{v11}[r][r]{1}%
\psfrag{v12}[r][r]{0}%
\psfrag{v13}[r][r]{90}%
\psfrag{v14}[r][r]{180}%
\psfrag{v15}[r][r]{0}%
\psfrag{v16}[r][r]{90}%
\psfrag{v17}[r][r]{180}%
%
\includegraphics[width=\columnwidth]{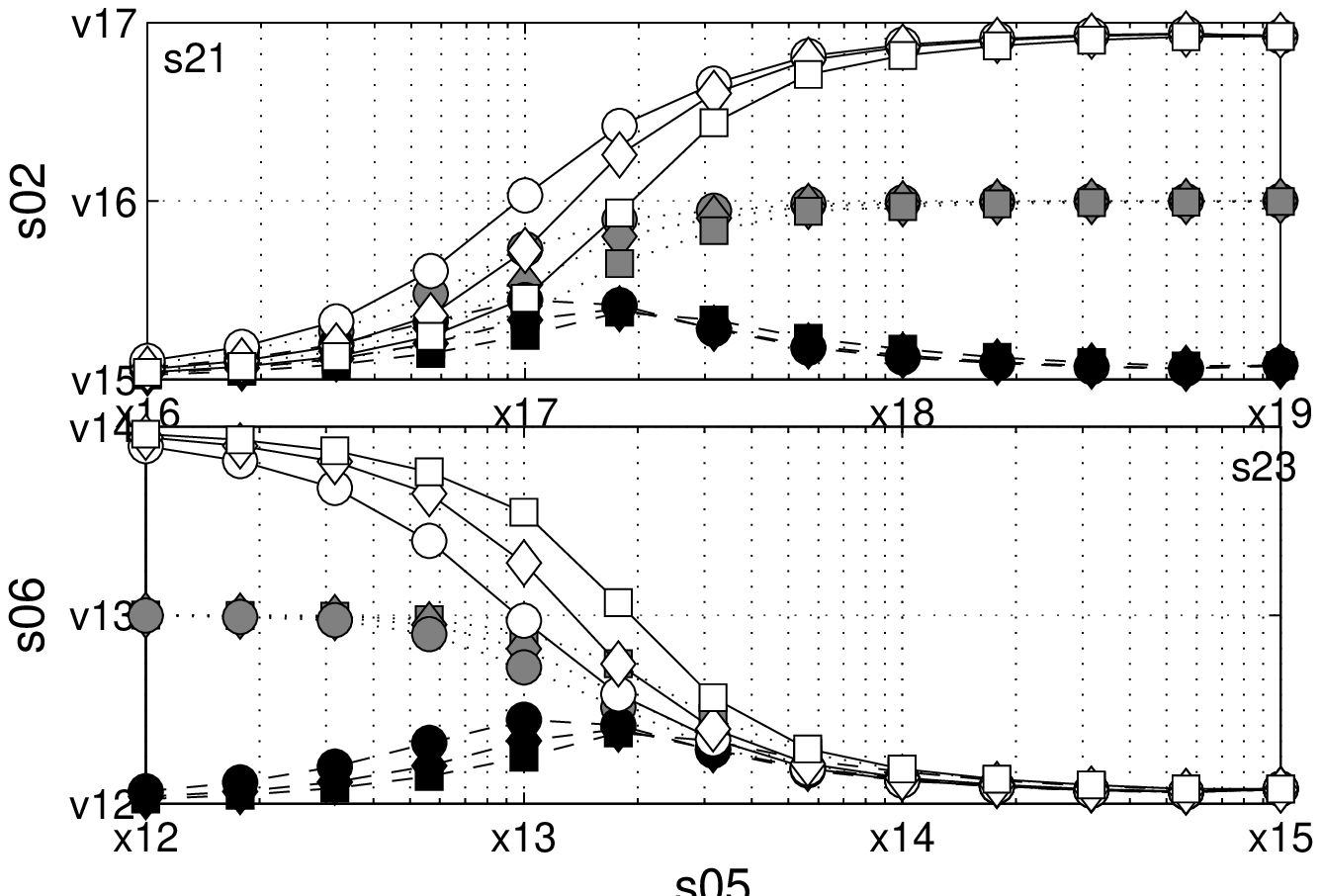}%
\end{psfrags}%
%

%% file: std_MpPDG.tex
%
%
\begin{psfrags}%
\psfragscanon%
%
\psfrag{s01}[t][t]{$D\;[ps/\sqrt{km}]$}%
\psfrag{s02}[b][b]{$STD(\theta_{\mathbf{M}\hat{P},\hat{\Gamma}})\;[^{\circ}]$}%
%
\psfrag{x01}[t][t]{0}%
\psfrag{x02}[t][t]{0.1}%
\psfrag{x03}[t][t]{0.2}%
\psfrag{x04}[t][t]{0.3}%
\psfrag{x05}[t][t]{0.4}%
\psfrag{x06}[t][t]{0.5}%
\psfrag{x07}[t][t]{0.6}%
\psfrag{x08}[t][t]{0.7}%
\psfrag{x09}[t][t]{0.8}%
\psfrag{x10}[t][t]{0.9}%
\psfrag{x11}[t][t]{1}%
\psfrag{x12}[t][t]{$10^{-4}$}%
\psfrag{x13}[t][t]{$10^{-3}$}%
\psfrag{x14}[t][t]{$10^{-2}$}%
\psfrag{x15}[t][t]{$10^{-1}$}%
%
\psfrag{v01}[r][r]{0}%
\psfrag{v02}[r][r]{0.1}%
\psfrag{v03}[r][r]{0.2}%
\psfrag{v04}[r][r]{0.3}%
\psfrag{v05}[r][r]{0.4}%
\psfrag{v06}[r][r]{0.5}%
\psfrag{v07}[r][r]{0.6}%
\psfrag{v08}[r][r]{0.7}%
\psfrag{v09}[r][r]{0.8}%
\psfrag{v10}[r][r]{0.9}%
\psfrag{v11}[r][r]{1}%
\psfrag{v12}[r][r]{0}%
\psfrag{v13}[r][r]{}%
\psfrag{v14}[r][r]{10}%
\psfrag{v15}[r][r]{}%
\psfrag{v16}[r][r]{20}%
\psfrag{v17}[r][r]{}%
\psfrag{v18}[r][r]{30}%
\psfrag{v19}[r][r]{}%
\psfrag{v20}[r][r]{40}%
\psfrag{v21}[r][r]{}%
%
\includegraphics[width=\columnwidth]{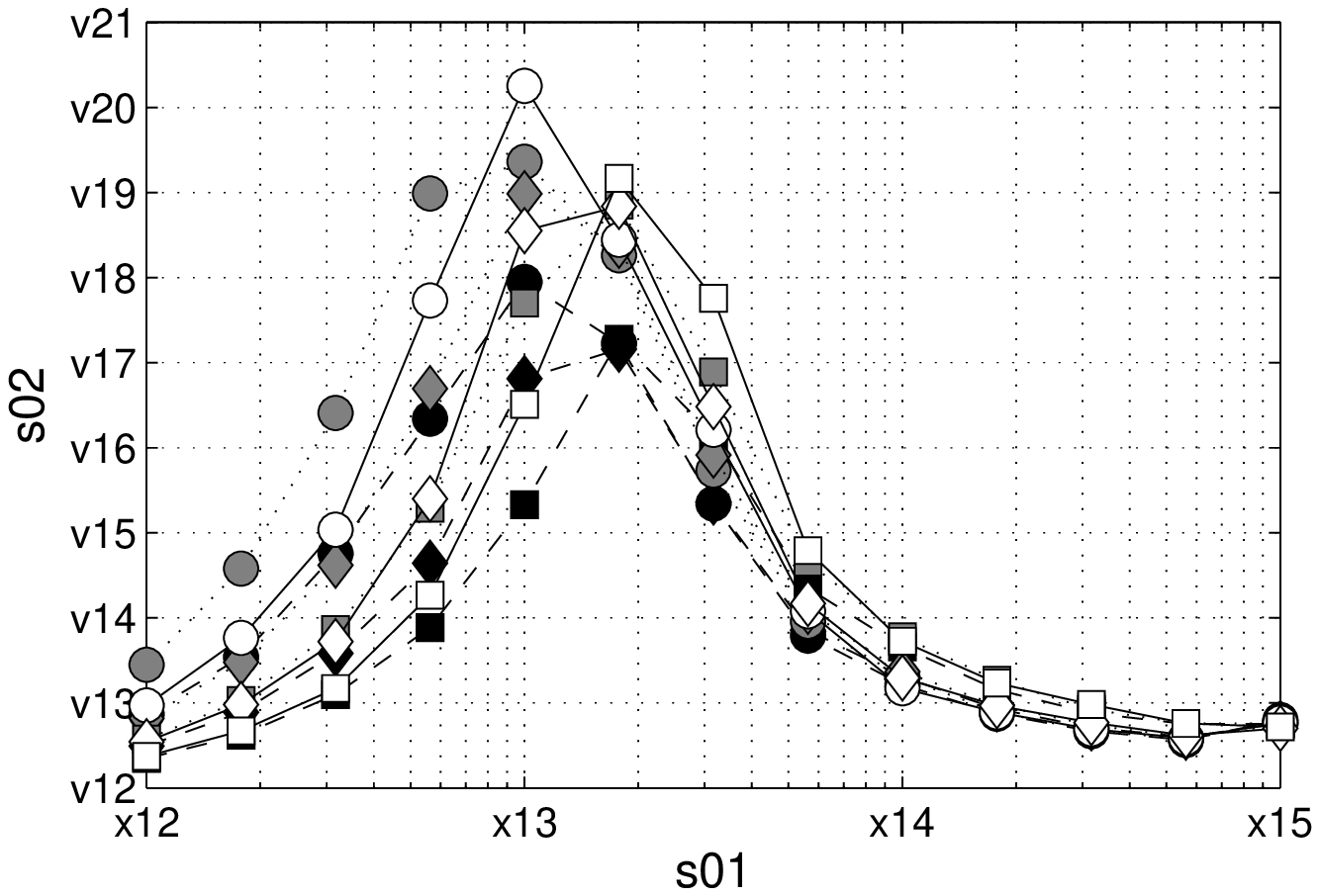}%
\end{psfrags}%
%